\documentclass[a4paper,11pt]{article}

\usepackage{jheppub}
\usepackage{amsmath,amssymb,amsthm}
\usepackage{graphicx}
\usepackage{booktabs}
\usepackage{orcidlink}

\usepackage{comment}

	\def\be{\begin{equation}}
		\def\ee{\end{equation}}
	\def\ba{\begin{eqnarray}}
		\def\ea{\end{eqnarray}}

\newcommand{\dd}{\mathrm{d}}
\newcommand{\lamL}{\lambda_L}
\newcommand{\Tind}{T_{\mathrm{ind}}}

\newcommand{\Oorb}{\Omega_{\mathrm{orb}}}

\newcommand{\rg}{r_{0}}

\newcommand{\tdot}{\dot t_{0}}

	\def \black hole { {\bar h} }

	\def\IR{\relax{\rm I\kern-.18em R}}
	\def\IL{\relax{\rm I\kern-.18em L}}
	
	\def\inv{^{\raise.15ex\hbox{${\scriptscriptstyle -}$}\kern-.05em 1}}

	\def\bea{\begin{eqnarray}}
		\def\eea{\end{eqnarray}}
	\definecolor{markcolor2}{rgb}{1,0,0}
	
	\definecolor{markcolor3}{rgb}{0,1,0}

    \def\dd{{\rm d}}

\usepackage{graphics,appendix,afterpage,makecell} 

\definecolor{oucrimsonred}{rgb}{0.6, 0.0, 0.0}
\definecolor{persianblue}{rgb}{0.11, 0.22, 0.73}
\definecolor{forestgreen}{rgb}{0.13,0.35,0.13}
\definecolor{lightgray}{rgb}{0.83, 0.83, 0.83}
\definecolor{cornellred}{rgb}{0.7, 0.11, 0.11}
\definecolor{navyblue}{rgb}{0.0, 0.0, 0.5}
\definecolor{amethyst}{rgb}{0.6, 0.4, 0.8}
\definecolor{yellow}{rgb}{1.0, 1.0, 0.0}
\definecolor{firebrick}{rgb}{0.7, 0.13, 0.13}
\definecolor{tangerineyellow}{rgb}{1.0, 0.8, 0.0}
\definecolor{deepfuchsia}{rgb}{0.76, 0.33, 0.76}
\definecolor{amber}{rgb}{1.0, 0.75, 0.0}
\definecolor{VioletRed4}{rgb}{0.55, 0.13, .32}
\definecolor{indiagreen}{rgb}{0.07, 0.53, 0.03}
\definecolor{VioletRed4}{rgb}{0.55, 0.13, .32}

\definecolor{oucrimsonred}{rgb}{0.6, 0.0, 0.0}
\newcommand\vertarrowbox[3][6ex]{%
  \begin{array}[t]{@{}c@{}} #2 \\
  \left\uparrow\vcenter{\hrule height #1}\right.\kern-\nulldelimiterspace\\
  \makebox[0pt]{\scriptsize#3}
  \end{array}%
}
\hypersetup{
     colorlinks   = true,
     citecolor    = violet,
     urlcolor     = violet,
     linkcolor    = violet}

\definecolor{verdechiaro}{rgb}{0.6,1,0.6}
\definecolor{giallochiaro}{rgb}{1,1,0.6}
\definecolor{bluscuro}{rgb}{0.15, 0.2, 0.9}
\definecolor{verdes}{rgb}{0.1, 0.5, 0.1}%
\definecolor{tangerineyellow}{rgb}{1.0, 0.8, 0.0}

\definecolor{americanrose}{rgb}{1.0, 0.01, 0.24}
\definecolor{cobalt}{rgb}{0.0, 0.28, 0.67}
\definecolor{brandeisblue}{rgb}{0.0, 0.44, 1.0}
\definecolor{mycolor}{rgb}{0.0, 0.0, 0.5}%navyblue
\definecolor{oxfordblue}{rgb}{0.0, 0.13, 0.28}
\definecolor{azure}{rgb}{0.0, 0.5, 1.0}
\definecolor{turquoiseblue}{rgb}{0.0, 1.0, 0.94}

\definecolor{verdes}{rgb}{0.1, 0.5, 0.1}%
\definecolor{cornellred}{rgb}{0.7, 0.11, 0.11}

\definecolor{VioletRed4}{rgb}{0.55, 0.13, .32}

\definecolor{rossocorsa}{rgb}{0.83, 0.0, 0.0}

\title{Thermal Origin of Black Hole Quasinormal Modes}

\author[a,b]{D.~Giataganas\orcidlink{0000-0003-2003-3902},}
\author[c,d]{G.F.~Giudice\orcidlink{0000-0002-0247-4096},} 
\author[e]{A.~Kehagias\orcidlink{0000-0001-6080-6215},}
\author[c,f]{F.~Quevedo\orcidlink{0000-0002-7810-3662},}
\author[g]{A.~Riotto\orcidlink{0000-0001-6948-0856}}

\affiliation[a]{Department of Physics, National Sun Yat-Sen University, Kaohsiung 80424, Taiwan
}
\affiliation[b]{Physics Division, National Center for Theoretical Sciences, Taipei 10617, Taiwan}

\affiliation[c]{New York University Abu Dhabi, Saadiyat Island, Abu Dhabi, UAE}
\affiliation[d]{CERN, Theoretical Physics Department, 1211 Geneva, Switzerland}
\affiliation[e]{Physics Division, National Technical University of Athens, Athens 15780, Greece}
\affiliation[f]{DAMTP, University of Cambridge, Wilberforce Road, Cambridge CB3 0WA, UK}
\affiliation[g]{Department of Theoretical Physics and Gravitational Wave Science Center,  \\
24 quai E. Ansermet, CH-1211 Geneva 4, Switzerland}

\rightline{CERN-TH-2026-189}

\abstract{
When a black hole rings after a merger, it emits gravitational waves at characteristic frequencies known as quasinormal modes (QNMs). In the eikonal limit, these modes are governed by the unstable circular light orbits that form the photon ring. In this work, we demonstrate that the ringing of a black hole has a precise thermal interpretation. A probe string propagating in the near-ring geometry acquires an induced Rindler horizon on its worldsheet, with a temperature set by the Lyapunov exponent of the photon ring. Out of this structure, the black hole QNMs emerge as thermal excitations, so  that the characteristic ringing of a black hole is the retarded response of a thermal system living on the photon ring. We explicitly derive the QNM spectrum from two complementary perspectives: microscopically, via unstable transverse worldsheet fluctuations, and macroscopically, through the pole structure of the causal response function of an open thermal quantum system.

}

\emailAdd{dimitrios.giataganas@gmail.com}
\emailAdd{gg3215@nyu.edu}
\emailAdd{kehagias@central.ntua.gr}
\emailAdd{fq2054@nyu.edu}
\emailAdd{Antonio.Riotto@unige.ch}
\begin{document}
\maketitle
\flushbottom

%=====================================================================
\section{Introduction}
\label{sec:introduction}
%=====================================================================

Black holes are among the simplest objects predicted by general relativity, yet their response to perturbations displays a remarkably rich structure. When a black hole is perturbed, as in the aftermath of a binary merger, it settles to its stationary state by emitting gravitational waves at a discrete set of complex frequencies, the quasinormal modes (QNMs)~\cite{Vishveshwara:1970zz,Press:1971wr,Chandrasekhar:1975zza,Kokkotas:1999bd,Nollert:1999ji,Berti:2009kk,Konoplya:2011qq,Berti:2025hly}. The real part of a QNM frequency determines the oscillation rate and sets the pitch of the ringing, while its imaginary part determines how rapidly the signal decays and sets the damping time. These frequencies depend only on the properties of the black hole geometry and therefore provide a direct way of probing the spacetime surrounding the compact object. Since the first direct detection of gravitational waves~\cite{LIGOScientific:2016aoc}, this ringdown phase has become a strong observational tool. Measuring the frequencies of the final black hole tests general relativity in the strong-field regime and probes the nature of the remnant, a program known as black hole spectroscopy~\cite{Berti:2005ys,Isi:2019aib}. Understanding why the QNM spectrum has the structure it does is therefore not only a conceptual question, but one directly tied to what current and future detectors measure.

In the short-wavelength, or eikonal, regime, the connection between QNMs and the motion of light near the black hole becomes particularly transparent. In this regime, a simple geometric picture underlies the spectrum. Light can orbit a black hole on unstable circular null geodesics (i.e.,~the photon ring) and eikonal QNMs correspond to wavepackets slowly leaking away from these orbits~\cite{Ferrari:1984zz,Mashhoon:1985cya,Cardoso:2008bp}. The real part of the frequency is set by the orbital frequency $\Oorb$ of the null orbit, while the damping is controlled by the Lyapunov exponent $\lamL$ measuring the instability rate of the orbit,
\begin{equation}
\label{eq:intro-eikonal}
  \omega_{mn} \simeq m\,\Oorb - i \left(n+\frac12\right)\lamL ,
  \qquad n=0,1,2,\ldots
\end{equation}
The same photon ring has become central to black hole imaging, where it controls the observed ring-like structure and its universal substructure~\cite{EventHorizonTelescope:2019dse,Gralla:2019xty,Johnson:2019ljv}. The photon ring is thus the common origin of how black holes look and how they sound.

More recently, this relation was understood in a more geometric and universal way in terms of the near-ring geometry \cite{Giataganas:2024hil}. However, several structural questions remain. Can we understand why the frequencies sit in the lower half of the complex plane, so that the ringing always decays and the overtones are organized in an evenly spaced ladder with spacing $\lamL$? Moreover, why does the tower start at the universal half-integer offset $n+1/2$, independently of the details of the black hole? 

These features are strongly reminiscent of thermal physics. The Lyapunov exponent plays a role analogous to surface gravity. The Maldacena--Shenker--Stanford (MSS) bound on quantum chaos states that the Lyapunov growth in a thermal quantum system cannot exceed $2\pi T$~\cite{Maldacena:2015waa} and is saturated by Rindler-like horizon dynamics. Moreover, conformal or thermal structures have been repeatedly encountered in near-photon-sphere physics~\cite{Hadar:2022xag}. Yet, these observations have largely remained analogies; the photon ring is not a horizon and no Hawking-like temperature is obviously associated with it.

In this paper, we argue that the thermal features of the eikonal QNM spectrum can be regarded as a consequence of an induced thermal field theory. Following ~\cite{Giataganas:2026ctn}, the construction uses a probe string as a diagnostic of the near-ring geometry. The relevant geometry is isolated through a Penrose limit along the unstable null orbit~\cite{Blau:2006ar,Giataganas:2024hil}, which retains the tidal information experienced by an observer moving along the photon ring~\cite{Blau:2006ar,Giataganas:2024hil,Kapec:2024lnr}. 
The probe string attached to the ring and stretched along the stable transverse direction then acquires an induced two-dimensional worldsheet metric that is precisely Rindler, with an induced surface gravity equal to the Lyapunov exponent saturating the MSS bound ~\cite{Giataganas:2026ctn}. By the standard logic of the Unruh effect and the Bisognano--Wichmann theorem~\cite{Unruh:1976db,Bisognano:1975ih,Bisognano:1976za,Crispino:2007eb}, the worldsheet state that is regular across this horizon, restricted to a single Rindler wedge, is a thermal Kubo--Martin--Schwinger (KMS) state at the induced temperature $\Tind \propto \lambda_L$, which is different from the black hole's Hawking temperature~\cite{Hawking:1975vcx}. Indeed, $\Tind$ is a worldsheet temperature attached to the photon ring rather than to the event horizon, and it is governed by orbital instability rather than by horizon surface gravity. In this sense, the Lyapunov exponent of the photon ring is translated into the temperature scale of a thermal quantum system.

Given this thermal system, the QNM spectrum emerges from two complementary directions. Microscopically, the unstable transverse fluctuation of the string possesses a zero mode governed by an inverted harmonic oscillator, whose outgoing (Gamow) resonances~\cite{Barton:1984ey} directly reproduce the damped tower of Eq.~\eqref{eq:intro-eikonal} in the frame co-rotating with the ring. 
Macroscopically, the same tower arises as the pole structure of the causal response of an open thermal system. The near-ring region is not closed, and excitations can leak to infinity or fall through the black hole horizon. Projecting onto the near-ring channel~\cite{Feshbach:1958nx,Feshbach:1962ut,Fano:1961zz} produces a retarded response whose analytic structure is fixed by general principles. Causality forces the resonance poles into the lower half of the complex-frequency plane, spectral positivity fixes the sign of the decay width, and KMS passivity~\cite{Pusz:1977hb} identifies this sign as thermal absorption. The worldsheet system can only damp the ringing, never amplify it. Finally, the universal half-integer offset  is fixed by the unitary half-density representation of Rindler boosts.

The resulting correspondence is then clear. The characteristic ringing of a black hole is the retarded response of a thermal system living on the photon ring, at a temperature set by the instability of light orbits. The microscopic worldsheet fluctuation identifies the dynamical origin of the resonance, while the thermal response explains its causal, dissipative, and universal structure.

The paper is organized as follows. In Section~\ref{sec:photon-ring-string} we present the near-ring Penrose limit and the probe string construction, derive the induced Rindler worldsheet and its temperature, and obtain the eikonal QNM spectrum from the outgoing resonances of the unstable transverse fluctuation. In Section~\ref{section3} we formulate the thermal properties of the worldsheet theory through the KMS condition and construct the real-time correlators relevant for the QNM response. In Section~\ref{section:causality} we explain, via causality and a Feshbach projection onto the escape channel, why the photon ring excitation appears as a decaying pole in the lower half-plane, with an absorptive width whose sign is dictated by KMS passivity. In Section~\ref{section:halfd} we determine the half-integer offset from the half-density representation of Rindler boosts acting on the projected escape coordinate, completing the derivation of Eq.~\eqref{eq:intro-eikonal}. We conclude in Section~\ref{section:conclusions}, and an appendix presents the effective two-point kernel realizing the half-density escape channel.

%=====================================================================
\section{QNMs from probe string worldsheet fluctuations}
\label{sec:photon-ring-string}
%=====================================================================

In this section, we recall the local geometry near the photon ring and the probe string embedding that gives rise to the thermal worldsheet theory \cite{Giataganas:2026ctn}.  The purpose is to make explicit which ingredients arise purely from the local photon ring geometry and which emerge only after quantization of the induced worldsheet theory.  

In the eikonal limit, black hole QNMs are controlled by the properties of unstable circular null geodesics, with the real part fixed by the orbital frequency and the damping controlled by the Lyapunov exponent of the orbit~\cite{Goebel:1972,Cardoso:2008bp,Yang:2012pj}.  Let us consider the equatorial circular null geodesic $\gamma$  
that generates the photon ring (in Boyer-Lindquist coordinates $(t,r,\theta,\phi)$),
\begin{equation}
r=r_0,\qquad \theta=\frac{\pi}{2},
\end{equation}
where $r_0$ is the photon ring radius. 
The orbital frequency measured with respect to the asymptotic time is
\begin{equation}
 \Oorb = \left.\frac{\dd\phi}{\dd t}\right|_{r=\rg}. \label{eq:Omegac}
\end{equation}
The near-ring geometry is obtained by taking the Penrose limit around the null
orbit $\gamma$~\cite{Blau:2006ar,Giataganas:2024hil}.  We now adopt Penrose-adapted null
coordinates $(u,v,x^1,x^2)$, where $u$ is the affine null coordinate along $\gamma$, $v$ is
the conjugate null coordinate, and $x^{1,2}$ are transverse Brinkmann coordinates.  In these
coordinates, the local metric takes the plane-wave form
\begin{equation}
\label{eq:ppwave}
\dd s^2  = 2\,\dd u\,\dd v+   A_{ij}(u)\,x^i x^j\,\dd u^2+   \dd x_i\,\dd x^i ,   \qquad i,j=1,2 .
\end{equation}
The tidal matrix $A_{ij}$ is obtained by projecting the Riemann tensor onto a parallel-propagated null frame along $\gamma$. The presence of a Killing–Yano tensor makes the analytic construction feasible and renders an otherwise challenging calculation nearly automatic for a broad class of stationary spacetimes \cite{Giataganas:2024hil}. The tidal matrix controls the transverse geodesic-deviation equation
\begin{equation}
\label{eq:deviation}
 \frac{\dd^2 x^i}{\dd u^2} =  A^i{}_{j}\,x^j .
\end{equation}
With this convention, a positive eigenvalue corresponds to an
unstable transverse direction since the corresponding mode obeys
an inverted oscillator equation. 

For an equatorial circular orbit in a four-dimensional vacuum geometry, the vacuum Einstein equations imply $\operatorname{tr}A=-R_{uu}=0$. For the stationary circular orbit considered here, the profile is constant in an adapted parallel-propagated eigenbasis. It may therefore be written as
\begin{equation}
\label{eq:tidal-diag}
A_{ij}= \mathrm{diag}\!\left(+A_{11},-A_{11}\right), \qquad   A_{11}>0 .
\end{equation}
Equivalently,
\begin{equation}
\label{eq:ppwave_diag}
 \dd s^2  =  2\,\dd u\,\dd v+  A_{11}\!\left[(x^1)^2-(x^2)^2\right]\dd u^2+  (\dd x^1)^2+(\dd x^2)^2 .
\end{equation}
With this convention and by looking at Eq. \eqref{eq:deviation}, $x^1$ is the unstable transverse
direction, while $x^2$ is the stable transverse direction.  The sign difference between the two Brinkmann eigenvalues is the local geometric imprint of the photon ring instability.  For the non-degenerate unstable photon orbit considered here, tracelessness implies a pair of opposite transverse eigenvalues. Thus, one transverse direction is unstable and the orthogonal direction is stable.

The Penrose limit retains only the local tidal information near the orbit.  It does not maintain the full parent spacetime null potential, but it does maintain its quadratic data at the photon ring peak, namely the unstable tidal eigenvalue $A_{11}$. Converting from affine time $u$ to physical asymptotic time,
\begin{equation}
\label{eq:t-redshift}
t=\tdot\,u,   \qquad  \tdot=\frac{\dd t}{\dd u}.
\end{equation}
Since $\dot t_0=dt/du$ is constant along the stationary circular orbit, converting Eq. \eqref{eq:deviation} from affine time $u$ to physical time $t$ gives the inverted harmonic oscillator for the unstable eigenmode 
\begin{equation}
\frac{d^2x^1}{dt^2}= \frac{A_{11}}{\tdot^{\,2}} x^1,
\end{equation}
from which the physical photon ring Lyapunov  instability exponent reads
\begin{equation}
\label{eq:A11-lambda}
\lamL^2=  \frac{A_{11}}{\tdot^{\,2}} .
\end{equation}
Thus, after converting the affine time used in the Penrose limit to the
physical time $t$, the plane-wave tidal frequency becomes precisely the
photon ring Lyapunov exponent~\cite{Cardoso:2008bp,Giataganas:2024hil}.  For example, for the equatorial Schwarzschild photon ring,
\begin{equation}
\label{eq:schw-A11}
\lamL^2=  \frac{A_{11}}{\tdot^{\,2}}= \frac{f(\rg)}{\rg^2}= \frac{1}{27M^2},  \qquad \rg=3M ,
\end{equation}
which also gives $\lamL^2=\Oorb^2$ in Schwarzschild, reproducing the known results.

\subsection{String background and induced Rindler worldsheet}
\label{sec:tidal-attached}

Following ~\cite{Giataganas:2026ctn}, we now introduce a probe string localized in the vicinity of the photon ring.
The string does not backreact on the geometry and, as we show, provides an extended diagnostic of the near-ring region. The probe string construction and the induced Rindler worldsheet reviewed in this subsection follow the near-ring analysis of Ref.~\cite{Giataganas:2026ctn}.

The probe string is described by embedding fields  $X^\mu = X^\mu(\xi^a)$,  $\xi^a = (\tau,\sigma)$, with Nambu--Goto action
\begin{equation}
S_{\rm NG} =  -T_s\int d^2\xi\,\sqrt{-\det G_{ab}},
\qquad
G_{ab}=  g_{\mu\nu}(X)\,\partial_a X^\mu\,\partial_b X^\nu .
\label{eq:NG}
\end{equation} 
We choose the static
gauge
\begin{equation}
 u = \frac{\tau}{\tdot},\qquad  x^2 = \sigma,
\label{eq:staticgauge}
\end{equation}
and allow the remaining embedding coordinates to fluctuate as
\begin{equation}
 v = v(\tau,\sigma),  \qquad x^1 = x^1(\tau,\sigma).
 \label{eq:fluct}
\end{equation}
With the pp-wave metric \eqref{eq:ppwave_diag}, the gauge-fixed Nambu--Goto
Lagrangian density becomes~\cite{Giataganas:2026ctn}
\begin{align}
\mathcal{L}_{\rm NG} = -T_s\Bigg\{
&\left(\frac{\partial_\sigma v}{\tdot}
  + \partial_\tau x^1\,\partial_\sigma x^1\right)^2  - \Bigg[\left(\frac{A_{11}}{\tdot^{\,2}}\right) \left((x^1)^2 - \sigma^2\right)
  \nonumber\\
&+ \frac{2\partial_\tau v}{\tdot}   +(\partial_\tau x^1)^2\Bigg]
  \left(1+(\partial_\sigma x^1)^2\right)\Bigg\}^{1/2}.
  \label{eq:LNG}
\end{align}
One immediately verifies that the straight string embedding
\begin{equation}
u = \frac{\tau}{\tdot},\qquad  x^2 = \sigma, \qquad  v(\tau,\sigma)=0, \qquad x^1(\tau,\sigma)=0,
\label{eq:background}
\end{equation}
solves the Nambu--Goto equations following from \eqref{eq:LNG}.  Pulling back
the plane-wave metric \eqref{eq:ppwave_diag} onto the  background \eqref{eq:background}  gives the worldsheet metric 
\begin{equation}
\label{eq:attached-rindler}
\dd s_{\rm ws}^2 = \frac{A_{22}}{\tdot^{\,2}}\sigma^2\,\dd\tau^2 + \dd\sigma^2
=   -\frac{A_{11}}{\tdot^{\,2}}\sigma^2\,\dd\tau^2 +   \dd\sigma^2 .
\end{equation}
Using \eqref{eq:A11-lambda}, this becomes
\begin{equation}
\label{eq:Rindler_ws}
\dd s_{\rm ws}^2 =-\kappa_{\rm ind}^2\sigma^2\,\dd\tau^2+ \dd\sigma^2,
\qquad
\kappa_{\rm ind}^2= \frac{A_{11}}{\tdot^{\,2}}= \lamL^2 .
\end{equation}
Thus the attached string has a genuine Rindler worldsheet. Its surface gravity is not the black hole surface gravity, but the induced worldsheet surface gravity equal to the photon ring Lyapunov exponent. The worldsheet horizon is located at $\sigma=0$, namely at the photon ring itself. The equality \eqref{eq:Rindler_ws} is purely geometric, and both the worldsheet acceleration scale and the photon ring instability are determined by the same Penrose tidal coefficient.

For the state that is regular across the two-sided worldsheet horizon, Euclidean regularity fixes the associated Rindler temperature. Setting $\tau=-i\tau_{\rm E}$, the induced metric becomes
\begin{equation}
\dd s^2_{E,{\rm ws}} =  \kappa_{\rm ind}^2\,\sigma^2\,\dd\tau_{\rm E}^2+  \dd\sigma^2 .
\label{eq:Euclidean_ws}
\end{equation}
This is the flat plane in polar coordinates, with radial coordinate $\sigma$
and angular coordinate $\kappa_{\rm ind}\tau_{\rm E}$.  Smoothness at $\sigma=0$
requires
$ \kappa_{\rm ind}\tau_{\rm E} \sim  \kappa_{\rm ind}\tau_{\rm E}+2\pi,$ 
or equivalently
\begin{equation}
\tau_{\rm E} \sim  \tau_{\rm E}+\beta_{\rm ind},  \qquad  \beta_{\rm ind}  =  \frac{2\pi}{\kappa_{\rm ind}} .
\label{eq:beta_ind}
\end{equation}
Euclidean regularity therefore fixes the inverse temperature associated with the regular worldsheet state to be
\begin{equation}
\label{eq:Tind}
\Tind=\frac{1}{\beta_{\rm ind}}= \frac{\kappa_{\rm ind}}{2\pi}=\frac{\lamL}{2\pi}.
\end{equation}

\subsection{String fluctuations}
\label{sec:tidal-fluc}

To isolate the orthogonal tidal dynamics, we now study the transverse string fluctuations in the Penrose-limit plane wave. We expand the
static gauge Nambu--Goto action~\eqref{eq:LNG} about the straight string background~\eqref{eq:background}, parametrising the fluctuation as
\begin{equation}
\label{eq:attached-fluc}
x^1(\tau,\sigma)= \epsilon\,\eta(\tau,\sigma), \qquad x^2=\sigma,  \qquad    u=\frac{\tau}{\tdot},  \qquad v=0 .
\end{equation}
Using $\kappa_{\rm ind}^2=A_{11}/\dot t_0^{\,2}$, the quadratic action for
the unstable transverse fluctuation is
\begin{equation}
S_2=\frac{T_s}{2}\int d\tau\,d\sigma\left[\frac{1}{\kappa_{\rm ind}\sigma}(\partial_\tau\eta)^2-\kappa_{\rm ind}\sigma
(\partial_\sigma\eta)^2+\frac{\kappa_{\rm ind}}{\sigma}\eta^2
\right],
\label{eq:attached-quadratic-action}
\end{equation}
where the constant background term and an overall factor of $\epsilon^2$ have been dropped.  The corresponding Euler--Lagrange equation is
\begin{equation}
\label{eq:penrose-transverse-string-eq}
\partial_\tau^2\eta-\kappa_{\rm ind}^2
\sigma\partial_\sigma\left(\sigma\partial_\sigma\eta\right)-\kappa_{\rm ind}^2\eta=0.
\end{equation}
Here the physical time is $t=\tau=\tdot\,u$ and from \eqref{eq:Rindler_ws} we know $\kappa_{\rm ind}=\lambda_L$. For the $\sigma$-independent mode, \eqref{eq:penrose-transverse-string-eq} reduces to
\begin{equation}
\label{eq:attached-zero-mode}
\ddot\eta_0 = \lamL^2\eta_0 ,
\end{equation}
the inverted oscillator relevant to the projected eikonal
escape channel, with the physical photon ring Lyapunov exponent. Thus, the zero
mode is the point-particle sector relevant for the standard eikonal QNM damping.
The role of the string background is to supply the induced Rindler worldsheet with $\kappa_{\rm ind}=\lambda_L$, while the orthogonal transverse Penrose fluctuation gives the unstable mode responsible for the QNM imaginary part.

It is useful to introduce the logarithmic Rindler radial coordinate $\rho=\ln\left(\sigma/\sigma_\ast \right),$ 
where $\sigma_\ast$ is an arbitrary reference scale. The fluctuation
equation then becomes
\begin{equation}
\left(\partial_\tau^2-\kappa_{\rm ind}^2\partial_\rho^2-\kappa_{\rm ind}^2\right)\eta=0.
\label{eq:eta-eq}
\end{equation}
We may therefore expand the fluctuation in Fourier modes of the
logarithmic radial coordinate,
\begin{equation}
\eta(\tau,\rho)
=\int_{-\infty}^{+\infty}\frac{dq}{2\pi}\,\eta_q(\tau)e^{iq\rho}=\int_{-\infty}^{+\infty}\frac{dq}{2\pi}\,\eta_q(\tau)\left(\frac{\sigma}{\sigma_\ast}\right)^{iq}.
\label{eq:attached-fourier}
\end{equation}
Inserting now \eqref{eq:attached-fourier} into \eqref{eq:eta-eq}, we find that 
the modes ${\eta}_q(\tau)$ satisfy the equation
\begin{equation}
\ddot{\eta}_q+\kappa_{\rm ind}^2\left(q^2-1\right)\eta_q=0.
\label{eq:attached-mode-eq}
\end{equation}
Equation~\eqref{eq:attached-mode-eq} shows that the modes with
$|q|<1$ are unstable, with the threshold $|q|=1$.  The fastest-growing
mode is the $q=0$ component, for which the fluctuation is independent
of $\sigma$ and the equation reduces to
\eqref{eq:attached-zero-mode}.  This generalized zero-momentum mode is
the inverted oscillator relevant to the projected eikonal escape
channel.  Its resonance condition is not a radial boundary condition
at $\sigma=0$, but the outgoing Gamow condition imposed on the
unstable coordinate $\eta_0$ in the next subsection.  The $q\neq0$
modes describe nonuniform extended worldsheet fluctuations and should
not be identified with the spacetime QNM overtone tower.

\subsection{Quasinormal frequencies from the co-rotating string fluctuation}
\label{sec:tidal-qnm}

The normalized zero-mode Hamiltonian is the inverted harmonic oscillator
\begin{equation}
\label{eq:io-H}
   H_{\rm ws}^{(0)}  = \frac12 p_\eta^2 - \frac12\lamL^2\eta_0^2 .
\end{equation}
The inverted oscillator has no normalizable bound-state spectrum with real energy. Its physically relevant discrete data are outgoing resonance poles, fixed by the outgoing Gamow condition, implemented by the complex rotation $\eta_0=e^{i\pi/4}y$ \cite{Barton:1984ey}. Under this rotation,
\begin{equation}
\label{eq:io-rotated-H}
H_{\rm ws}^{(0)} \longrightarrow -i \left(\frac12 p_y^2+\frac12\lamL^2 y^2\right) \equiv-i\,H_{\rm ho}.
\end{equation}
Since the ordinary harmonic oscillator has eigenvalues 
\begin{equation}\label{eq:eigen}
H_{\rm ho}\psi_n=\lamL(n+\tfrac12)\psi_n,
\end{equation}
the inverted oscillator possesses a discrete set of outgoing resonance poles (Gamow poles)
\begin{equation}
\label{eq:io-gamow-poles}
E_n = -i\,\lamL\left(n+\frac12\right),  \qquad   n=0,1,2,\ldots .
\end{equation}
which decay as $e^{-iE_n\tau}$ in time. This is the spectrum of the fluctuation in the frame
co-moving with the background string, co-rotating with  $\Oorb$ \eqref{eq:Omegac}, as
\begin{equation}
t=\tau, \qquad \phi=\Oorb\tau .
\end{equation}
Therefore, a spacetime mode with phase  $ e^{-i\omega t+im\phi}$ for integer $m$ pulls back to the worldsheet as $e^{-i(\omega-m\Oorb)\tau}.$
Thus, the frequency entering the local tidal fluctuation problem is the co-rotating
worldsheet frequency
\begin{equation}
\label{eq:corot-frequency}
\omega^{\rm cr}=\omega-m\Oorb .
\end{equation}
Since the zero-mode Hamiltonian generates evolution with respect to the worldsheet time $\tau$, its resonance energy is identified with the frequency conjugate to $\tau$.  Using \eqref{eq:io-gamow-poles}, the outgoing Gamow poles of the transverse inverted oscillator are therefore
\begin{equation}
\label{eq:nu-poles}
\omega^{\rm cr}_n = -i\left(n+\frac12\right)\lamL .
\end{equation}
Undoing the pullback relation \eqref{eq:corot-frequency}, one obtains the
equatorial eikonal QNM frequencies
\begin{equation}
\label{eq:qnm}
\omega_{mn}=m\Oorb- i\left(n+\frac12\right)\lamL,
\end{equation}
for equatorial modes in the large-$m$ eikonal regime. Thus, both parts of the eikonal QNM frequency arise from the same string construction, where the real part is supplied by the longitudinal orbital motion of the background string, while the imaginary part is supplied by the transverse tidal
inverted oscillator.

In summary, the Rindler structure  arose from the string embedding \eqref{eq:background}, giving the induced metric \eqref{eq:Rindler_ws} with a Rindler horizon at $\sigma=0$, i.e., at the photon ring.  Euclidean regularity of the worldsheet state fixes the induced Unruh temperature \eqref{eq:Tind}. The unstable transverse fluctuation is the orthogonal field $x^1=\eta$, whose $\sigma$-independent mode is an inverted
oscillator obeying \eqref{eq:attached-zero-mode},  whose outgoing Gamow poles reproduce the eikonal QNM spectrum of \eqref {eq:qnm}.

%%=====================================================================
\section{KMS thermality and real-time worldsheet correlators}
\label{section3}
%%=====================================================================
In this section, we formulate the thermal properties of the worldsheet theory in terms of the KMS condition and introduce the real-time correlators needed to describe the QNM response. It is important to distinguish the local inverted-oscillator generator
$H_{\rm ws}^{(0)}$ from the Hamiltonian used to formulate the KMS
state $H_R$.  The complex frequencies in \eqref{eq:io-gamow-poles} are resonance poles obtained after imposing the outgoing Gamow condition. They are not eigenvalues in the normalizable spectrum of the self-adjoint Hamiltonian and therefore are not used to define a thermal density operator.

Throughout this section and
Section~\ref{section:causality}, $H_R$ denotes the self-adjoint
generator of Rindler-time translations in an assumed complete,
unprojected effective system.  The inverted oscillator Hamiltonian 
 $H_{\rm ws}^{(0)}$ is interpreted as the local near-ring normal form of
the projected resonance dynamics of this complete system. We assume that this system admits a regular two-sided,
boost-covariant completion to which the
Bisognano--Wichmann theorem applies. The Rindler coordinates are related to locally inertial worldsheet light-cone coordinates by
\begin{equation}
X^+ = \sigma\,e^{\kappa_{\rm ind}\tau},\qquad
X^- = -\sigma\,e^{-\kappa_{\rm ind}\tau},\qquad
ds^2_{\rm ws} = -dX^+dX^-.
\label{eq:lightcone}
\end{equation}
These coordinates reproduce the worldsheet metric  \eqref{eq:Rindler_ws} and, for $\sigma>0$,  they cover the right Rindler wedge. A shift in worldsheet time acts as a boost:
\begin{equation}
\tau\to\tau+\Delta\tau,\qquad
X^+\to e^{\kappa_{\rm ind}\Delta\tau}X^+,\qquad
X^-\to e^{-\kappa_{\rm ind}\Delta\tau}X^-.
\label{eq:boost}
\end{equation}
Denoting the dimensionless geometric boost generator of the regular
completion by $K_{\rm BW}$, comparison of the boost flow
\eqref{eq:boost} with translations in $\tau$ gives
\begin{equation}
H_R=\kappa_{\rm ind}K_{\rm BW},\qquad H_R^\dagger=H_R.
\label{eq:HR_KBW}
\end{equation}
Under the regular-completion assumption stated above, the
Bisognano--Wichmann theorem
\cite{Bisognano:1975ih,Bisognano:1976za} identifies the modular flow of
the Minkowski vacuum restricted to either Rindler wedge with this
geometric boost flow.  The restricted state is therefore a KMS state
with respect to $H_R$ at
\begin{equation}
\beta_{\rm ind}=\frac{2\pi}{\kappa_{\rm ind}}, \qquad \beta_{\rm ind}H_R=2\pi K_{\rm BW}.
\label{eq:BW_temperature}
\end{equation}

In continuum quantum field theory, the wedge state is most precisely
defined through the algebraic KMS condition rather than by a
trace-class density matrix, and with an implicit regulator, it may
 be written formally as
\begin{equation}
\rho_{\rm ws}=\frac{e^{-\beta_{\rm ind}H_R}}{Z_R}=\frac{e^{-2\pi K_{\rm BW}}}{Z_R},
\qquad
Z_R=\operatorname{Tr}e^{-\beta_{\rm ind}H_R}.
\label{eq:rho_ws}
\end{equation}

Since $H_R$ generates translations in $\tau$, a worldsheet operator in the Heisenberg picture evolves as 
\begin{equation}
  \mathcal{O}(\tau) = e^{iH_R\tau}\,\mathcal{O}(0)\,e^{-iH_R\tau}.
  \label{eq:Heisenberg}
\end{equation}

The regular state restricted to a single worldsheet Rindler wedge is therefore thermal with respect to the Rindler-time $\tau$ evolution.  For bosonic worldsheet operators, the associated thermal correlators satisfy the KMS condition
\begin{equation}
  \langle A(\tau)B(0)\rangle = \langle B(0)A(\tau+i\beta_{\rm ind})\rangle.
  \label{eq:KMS}
\end{equation}

In order to study the frequency spectrum of the system, we  use the Fourier-transform convention
\begin{equation}\label{eq:fourier}
{\tilde G}(\omega)=\int_{-\infty}^{+\infty}d\tau\,e^{i\omega\tau}G(\tau),
\end{equation}
where, in this section and in Section~\ref{section:causality}, the frequency
$\omega$ is conjugate to the worldsheet time $\tau$ and therefore
corresponds to the co-rotating frequency $\omega^{\rm cr}$ defined in 
\eqref{eq:corot-frequency}.  The asymptotic orbital contribution is
restored only when returning to the spacetime frequency.

The two thermal Wightman functions are defined in the standard real-time notation ~\cite{Bellac:2011kqa} by
\begin{align}
G^>(\tau) &=\langle\mathcal{O}(\tau)\mathcal{O}(0)\rangle_{\beta_{\rm ind}}= \frac{1}{Z_R}\mathrm{Tr}\!\left[e^{-\beta_{\rm ind}H_R}\,\mathcal{O}(\tau)\,\mathcal{O}(0)\right],\\
\quad
G^<(\tau) &=\langle\mathcal{O}(0)\mathcal{O}(\tau)\rangle_{\beta_{\rm ind}}= \frac{1}{Z_R}\mathrm{Tr}\!\left[e^{-\beta_{\rm ind}H_R}\,\mathcal{O}(0)\,\mathcal{O}(\tau)\right].
  \label{eq:Wightman}
\end{align}
These functions are not time-ordered correlators as they differ in the ordering of the two operator insertions. These orderings are generally inequivalent because operators at different times need not commute. 

The function $G^>(\tau)$ measures the correlation
between an operator inserted at initial time $\tau=0$ and the same operator measured later at $\tau>0$.   The function $G^<(\tau)$ contains the same two insertions in the opposite operator order.

In frequency space, ${\tilde G}^>(\omega)$ measures the spectral weight for processes in which the worldsheet system absorbs Rindler energy~$\omega$, while ${\tilde G}^<(\omega)$ measures the corresponding reversed emission process.  The KMS condition relates these two processes by the Boltzmann factor.  Thus, $G^>$ and $G^<$ are the basic real-time correlators from which the spectral density, the retarded response, and the fluctuation--dissipation relation are
constructed.
 
In frequency space the sign of $\omega$ labels the direction of energy transfer.  For $\omega>0$, ${\tilde G}^>(\omega)$ gives the spectral weight for transitions to states of higher worldsheet Rindler energy, and therefore corresponds to absorption by the worldsheet thermal system.  On the other hand, ${\tilde G}^<(\omega)$ is related by KMS to the reverse process, namely emission from the thermal system.  The relation between the two is detailed balance. By using \eqref{eq:Heisenberg} and the cyclicity of the trace, we find that
\begin{align}
G^>(\tau-i\beta_{\rm ind})= &\frac{1}{Z_R}\mathrm{Tr}\!\left[e^{-\beta_{\rm ind}H_R}\,\mathcal{O}(\tau-i\beta_{\rm ind})\,\mathcal{O}(0)\right]
\nonumber \\
=& \frac{1}{Z_R}\mathrm{Tr}\!\left[\mathcal{O}(\tau)\,e^{-\beta_{\rm ind}H_R}\,\mathcal{O}(0)\right]= G^<(\tau).
\label{eq:KMS_derive}
\end{align}
Equation~\eqref{eq:KMS_derive} is the bosonic  KMS condition~\cite{Kubo:1957mj,Martin:1959jp,Haag:1967sg}.   

A state satisfying the KMS condition with respect to the Rindler Hamiltonian $H_R$ is thermal and determines the thermal analyticity properties of its correlation functions. Equivalently, the KMS condition is
the Lorentzian counterpart of the Euclidean thermal circle.  After continuing to Euclidean
Rindler time, bosonic correlators are periodic with period $\beta$, $G_{\rm E}(\tau_{\rm E}+\beta_{\rm ind}) = G_{\rm E}(\tau_{\rm E}),$ while fermionic correlators are anti-periodic. Thus, the Euclidean thermal circle derived from regularity is equivalently encoded as the KMS periodicity of Euclidean correlation functions.

The KMS relation \eqref{eq:KMS_derive}  gives  the condition $G^>(\tau) = G^<(\tau+i\beta_{\rm ind}),$ whereas by Fourier transforming \eqref{eq:KMS_derive} we get
\begin{equation}
{\tilde G}^<(\omega) = e^{-\beta_{\rm ind}\omega}\,{\tilde G}^>(\omega).
\label{eq:KMS_fourier}
\end{equation}
Now, for the spectral density defined by
\begin{equation}
\rho_\mathcal{O}(\omega) \equiv {\tilde G}^>(\omega) - {\tilde G}^<(\omega)=\int_{-\infty}^{+\infty}d\tau\,
e^{i\omega\tau} \left\langle[\mathcal O(\tau),\mathcal O(0)]\right\rangle_{\beta_{\rm ind}},
\label{eq:spectral} 
\end{equation}
equation \eqref{eq:KMS_fourier} implies that
\begin{equation}
{\tilde G}^>(\omega) = \frac{\rho_\mathcal{O}(\omega)}{1-e^{-\beta_{\rm ind}\omega}},\qquad
{\tilde G}^<(\omega) = \frac{e^{-\beta_{\rm ind}\omega}\,\rho_\mathcal{O}(\omega)}{1-e^{-\beta_{\rm ind}\omega}},
\label{eq:spectral_decomp}
\end{equation}
where $\rho_{\cal O}$ is equal to the Fourier transform of the thermal commutator.
Thus,  the KMS condition  fixes the detailed-balance relation between absorption and emission in the worldsheet
response.
 
For a stationary equilibrium state, the KMS condition implies the fluctuation–dissipation relation.  From the
Wightman functions $G^>$ and $G^<$ we can form the retarded and the symmetrised correlator
 $G^R$ and $G^{\rm sym}$, respectively, as
\begin{equation}
G^R(\tau) = -i\theta(\tau)\!\left[G^>(\tau)-G^<(\tau)\right],\qquad
G^{\rm sym}(\tau) = \tfrac{1}{2}\!\left[G^>(\tau)+G^<(\tau)\right].
\label{eq:GR_Gsym}
\end{equation}
Thus, $G^R$ probes the commutator part of the thermal response, whereas $G^{\rm sym}$ probes the
anticommutator part and measures the size of the equilibrium fluctuations.  KMS relates the two Wightman functions by \eqref{eq:KMS_fourier}, 
and therefore fixes the relation between these two real-time responses. Using \eqref{eq:spectral_decomp}, we find
\begin{equation}
{\tilde G}^{\rm sym}(\omega) = \frac{1}{2}\coth\!\!\left(\frac{\beta_{\rm ind}\omega}{2}\right)\rho_\mathcal{O}(\omega).
\label{eq:Gsym}
\end{equation}
With the retarded and Fourier conventions adopted above, it satisfies
\begin{equation}
\rho_\mathcal{O}(\omega) = -2\,\mathrm{Im}\,{\tilde G}^R(\omega).
\label{eq:spectral_Im}
\end{equation}
Substituting this result into  \eqref{eq:Gsym} we obtain
\begin{equation}
{\tilde G}^{\rm sym}(\omega) = -\coth\!\!\left(\frac{\beta_{\rm ind}\omega}{2}\right)\mathrm{Im}\,{\tilde G}^R(\omega).
\label{eq:FDT}
\end{equation}
This is the fluctuation--dissipation relation.  In the worldsheet problem, it means that the
thermal fluctuations seen by a Rindler observer and the dissipative part of the near-ring
response are related by KMS. It provides a direct connection between fluctuations of the thermal worldsheet theory and dissipation in the near-ring subsystem,
\begin{equation}
\text{worldsheet fluctuations}
\quad\longleftrightarrow\quad
\text{near-ring dissipation}.
\label{eq:ws-fluctuation-dissipation}
\end{equation}
The KMS condition implies that thermal fluctuations experienced by a Rindler observer on the string worldsheet and the absorptive response of the near-ring channel are not independent. Once the worldsheet temperature and the spectral response are specified, the strength of the equilibrium worldsheet fluctuations is fixed by the dissipative part of the response. In this sense, the projected channel whose retarded response describes QNM leakage also exhibits equilibrium fluctuations governed by the same spectral density and induced temperature $T_{\rm ind}$ given by \eqref{eq:Tind}.

Finally, KMS provides analyticity for thermal Wightman functions.  The functions $G^>(\tau)$ and $G^<(\tau)$ are boundary values of a single correlator that is analytic in a strip of imaginary time of width~$\beta_{\rm ind}$.  The two boundaries of this strip are related by the KMS condition, where shifting one operator by $i\beta_{\rm ind}$ in imaginary time reverses the operator ordering,
up to the statistics sign.  This is the precise thermal analyticity property used in finite-temperature QFT.  It is also the analyticity input that enters the derivation of the thermal chaos bound~\cite{Maldacena:2015waa}.

For the photon ring problem, the physical meaning is the following. Once the probe string reveals the induced Rindler geometry and the worldsheet field is placed in the regular state, near-ring observables are described by KMS correlators at $T_{\rm ind}$ satisfying \eqref{eq:KMS}. QNMs are poles of retarded response functions subject to the appropriate outgoing boundary conditions ~\cite{Nollert:1999ji,Kokkotas:1999bd}. Causality requires the retarded Green function to be analytic in the upper half-plane. For a stable dissipative channel with outgoing boundary conditions, its resonance poles therefore lie in the lower half-plane. The evenly spaced pole structure considered below follows only after specifying the projected escape-channel correlator.  In the eikonal limit, these poles are controlled by photon ring geodesic data~\cite{Ferrari:1984zz,Mashhoon:1985cya}.

%%%=====================================================================
\section{Causality, dissipation, and QNM leakage}
%%%=====================================================================
\label{section:causality}
 
Here we explain why the photon ring excitation appears as a decaying QNM pole rather than as a normal mode with a real frequency. 

\subsection{Causality}
\noindent
Causality in the QNM problem is most naturally formulated in terms of a  linear response. A QNM is not a normal mode of a closed system, but a resonance seen in the response of the near-ring region after it has been perturbed.  The physical response must be causal, i.e., a perturbation applied at worldsheet time $\tau=0$ cannot affect the expectation value of a near-ring observable at earlier times $\tau<0$.  In quantum field theory, this causal response is encoded by the retarded Green function.  Thus, if $\mathcal{O}_{\rm esc}(\tau)$  denotes an effective projected worldsheet operator probing the near-ring escape channel, the relevant response function is
\begin{equation}
G^R_{\rm esc}(\tau) = -i\theta(\tau)\,\langle[\mathcal{O}_{\rm esc}(\tau),\mathcal{O}_{\rm esc}(0)]\rangle_{\beta_{\rm ind}}.
\label{eq:GResc}
\end{equation}
The step function is the causal input that states that the response vanishes before the source
is applied.  The commutator is the quantum-mechanical linear-response kernel. If the worldsheet
Hamiltonian is perturbed by a source coupled to $\mathcal{O}_{\rm esc}$, then the induced
change in the expectation value of $\mathcal{O}_{\rm esc}$ is controlled by $G^R_{\rm esc}$. This is the Kubo linear-response relation. Therefore, the QNM frequencies of the near-ring channel are identified with the poles of this
retarded response function.
 
The causal property in Eq.~\eqref{eq:GResc} immediately imposes an analyticity
condition on ${\tilde G}^R_{\rm esc}(\omega)$.  With the Fourier convention \eqref{eq:fourier}
we have
\begin{equation}
{\tilde G}^R_{\rm esc}(\omega) = -i\int_0^{+\infty}d\tau\,e^{i\omega\tau}\langle[\mathcal{O}_{\rm esc}(\tau),\mathcal{O}_{\rm esc}(0)]\rangle_{\beta_{\rm ind}}.
\label{eq:GResc_fourier}
\end{equation}
Then, for complex frequency $\omega=\omega_R+i\omega_I$, the Fourier factor behaves as $e^{i\omega\tau} = e^{i\omega_R\tau}e^{-\omega_I\tau}.$ Therefore, for $\omega_I>0$, the integral is exponentially damped at large positive $\tau$. Under the usual stability and boundedness assumptions on thermal correlators, this implies that 
${\tilde G}^R_{\rm esc}(\omega)$ is analytic for $\mathrm{Im}\,\omega>0$.

Thus a causal retarded response cannot have physical poles in the upper half of the complex-frequency plane. Causality gives upper half-plane analyticity of the retarded response. Dissipation of the near-ring channel produces resonance poles.  Therefore the poles of a stable causal leaking channel must lie in the lower half-plane.

 \subsection{Feshbach projection and positive width}
 \noindent
It is important to distinguish the causal location of a resonance pole
from the physical origin of its width.  Causality, together with
stability, gives upper-half-plane analyticity of the retarded response,
but it does not by itself generate a nonzero imaginary part.  The width
arises when the near-ring channel is treated as an open subsystem
coupled to complementary propagation channels.
We describe this structure by applying a Feshbach projection
\cite{Feshbach:1958nx,Feshbach:1962ut,Fano:1961zz} to the same
self-adjoint Hamiltonian $H_R$ that generates the KMS evolution in
Section~\ref{section3}.  The projectors $P$ and $Q$ are effective
projectors adapted to the near-ring response. They are not obtained
here from an explicit mode-by-mode decomposition of the complete
gauge-fixed Nambu--Goto theory.  The $P$ sector retains the near-ring
escape channel probed by $\mathcal O_{\rm esc}$, while $Q=1-P$ contains
the complementary channels eliminated from the projected retarded
response.
 
We therefore introduce the effective decomposition
$\mathcal H=\mathcal H_P\oplus\mathcal H_Q$.  The projector $P$ selects the near-ring sector, while $Q=1-P$ selects the complementary propagation channels through which an excitation can leave the
near-ring region.  Upon embedding the local Penrose description into
the complete black-hole scattering problem, these channels are
expected to correspond to flux propagating toward asymptotic infinity
or into the black-hole horizon.  The projectors $P$ and $Q$ are
orthogonal and satisfy
\begin{equation}
P^\dagger=P,\qquad Q^\dagger=Q,\qquad P^2=P,\qquad Q^2=Q,\qquad PQ=QP=0.
\label{eq:projectors}
\end{equation}
The $P$ sector is the sector of states localised near the photon ring, while the $Q$ sector contains the
degrees of freedom into which the near-ring excitation can leak, associated
with flux escaping away from the near-ring region or being absorbed through the horizon.  The
block components of the same worldsheet Hamiltonian $H_R$ are
\begin{equation}
H_{PP} = PH_RP,\quad H_{QQ} = QH_RQ,\quad  H_{PQ} = PH_RQ,\quad H_{QP} = QH_RP = H_{PQ}^\dagger.
\label{eq:Hblocks}
\end{equation}
Thus $H_{PP}$ is the Hamiltonian restricted to the near-ring sector, $H_{QQ}$ is the
Hamiltonian restricted to the complementary leakage channels, and $H_{PQ}$, $H_{QP}$ couple the two
sectors.  These are the block components of the same
self-adjoint Rindler Hamiltonian~$H_R$.

Let $|\Psi\rangle=|\psi_P\rangle+|\psi_Q\rangle$ be the state generated by a source $|\phi_P\rangle$ lying entirely in the P sector.
The projected retarded Green function in the near-ring sector is the $PP$ block of the full
resolvent
\begin{equation}
{\tilde G}^R_P(\omega) = P\,\frac{1}{\omega-H_R+i0^+}\,P.
\label{eq:GRP}
\end{equation}
Then the $P$-sector equation gives
\begin{equation}
{\tilde G}^R_P(\omega) = \frac{1}{\omega - H_{PP} - \Sigma^R(\omega)},
\label{eq:Feshbach}
\end{equation}
where the retarded self-energy is
\begin{equation}
\Sigma^R(\omega) = H_{PQ}\,\frac{1}{\omega-H_{QQ}+i0^+}\,H_{QP}.
\label{eq:SelfEnergy}
\end{equation}
We may therefore define the energy-dependent retarded effective Hamiltonian acting
in the projected sector by
\begin{equation}
H_{\rm eff}^R(\omega)\equiv H_{PP}+\Sigma^R(\omega),
\label{eq:HeffR}
\end{equation}
so that
\begin{equation}
{\tilde G}_P^R(\omega)=\frac{1}{\omega-H_{\rm eff}^R(\omega)}.
\label{eq:GRP_Heff}
\end{equation}

Although $\mathcal O_{\rm esc}$ probes the $P$ sector, its correlation
functions are evaluated in the KMS state of the complete system, and
its Heisenberg evolution is generated by the self-adjoint Hamiltonian $H_R$.  By contrast, $H_{\rm eff}^R(\omega)$ describes only the projected retarded resolvent, it is  generically non-Hermitian, and no
independent KMS state is assigned to this effective operator.  The
standard retarded prescription~\cite{Srednicki_2007} then gives
\begin{equation}
\frac{1}{\omega-H_{QQ}+i0^+} = \mathrm{PV}\frac{1}{\omega-H_{QQ}} - i\pi\delta(\omega-H_{QQ}),
\label{eq:retarded_prescription}
\end{equation}
where $\mathrm{PV}$ denotes the principal value.  Therefore the retarded self energy develops a real dispersive part and an imaginary  absorptive part
\begin{equation}
\Sigma^R(\omega) = \Delta(\omega) - \frac{i}{2}\Gamma(\omega),
\label{eq:SelfEnergy_decomp}
\end{equation}
with
\begin{equation}
\Delta(\omega) = H_{PQ}\,\mathrm{PV}\frac{1}{\omega-H_{QQ}}\,H_{QP},\qquad
\Gamma(\omega) = 2\pi\,H_{PQ}\,\delta(\omega-H_{QQ})\,H_{QP}.
\label{eq:Gamma}
\end{equation}
Thus the Feshbach blocks are the near-ring and exterior-channel blocks of the same Rindler
Hamiltonian $H_R=\kappa_{\rm ind}K_{\rm BW}$ that defines the KMS state.  The imaginary part of the projected self-energy appears because the P sector is treated as an open subsystem coupled to Q.  If the full $P\oplus Q$ system is
kept, the evolution generated by $H_R$ remains unitary.  Dissipation appears only in the
projected near-ring response.
 
The positivity of the width $\Gamma(\omega)$ follows directly from the spectral theorem.  For
any state $|\psi_P\rangle$ in the near-ring subspace, using \eqref{eq:Gamma} we have
\begin{equation}
\langle\psi_P|\Gamma(\omega)|\psi_P\rangle  =2\pi\langle\psi_P|H_{PQ}\,\delta(\omega-H_{QQ})\,H_{QP}|\psi_P\rangle,
\label{eq:Gamma_pos1}
\end{equation}
Defining $|\chi_Q\rangle=H_{QP}|\psi_P\rangle$ which lies in the $Q$ sector, this becomes
\begin{equation}
\langle\psi_P|\Gamma(\omega)|\psi_P\rangle= 2\pi\langle\chi_Q|\delta(\omega-H_{QQ})|\chi_Q\rangle.
\label{eq:Gamma_pos2}
\end{equation}
Since the full Hamiltonian is self-adjoint and by using that  $H_{QQ}|\alpha\rangle=E_\alpha|\alpha\rangle$, we see that $H_{QQ}$ has a positive spectral measure 
\begin{equation}
\langle\chi_Q|\delta(\omega-H_{QQ})|\chi_Q\rangle = \sum_{\alpha\in Q}|\langle\alpha|\chi_Q\rangle|^2\,\delta(\omega-E_\alpha) \geq 0.
\label{eq:spectral_positive}
\end{equation}
 Hence
\begin{equation}
\Gamma(\omega) \geq 0
\label{eq:Gamma_positive}
\end{equation}
is positive semidefinite. Physically, matrix elements of $\Gamma(\omega)$ give the golden-rule transition rates from the near-ring sector into the open Q-sector channels. The full $P\oplus Q$ evolution remains unitary, while the reduced P-sector dynamics becomes dissipative after the Q-sector degrees of freedom are integrated out. For illustration, suppose that the projector $P$ is chosen so that $H_{PP}$ contains an isolated near-ring state $|r\rangle$ with
\begin{equation}
H_{PP}|r\rangle=\Omega|r\rangle, \qquad \langle r|r\rangle=1.
\end{equation}
In the single-pole and weak-width approximation, the corresponding pole
of \eqref{eq:Feshbach} is
\begin{equation}
\omega_{\rm pole} = \Omega+\Delta_r(\Omega)-\frac{i}{2}\Gamma_r(\Omega),
\label{eq:pole}
\end{equation}
where
\begin{equation}
\Delta_r(\omega)=\langle r|\Delta(\omega)|r\rangle,
\qquad
\Gamma_r(\omega)=\langle r|\Gamma(\omega)|r\rangle.
\end{equation}
The negative imaginary part is the mathematical expression of leakage
from the projected near-ring sector.
 
The positivity of $\Gamma(\omega)$ also has a direct thermal interpretation.  This positivity follows independently from the spectral theorem. In a KMS state it is consistent with, and receives a thermal interpretation from, passivity~\cite{Pusz:1977hb}. Using \eqref{eq:KMS_fourier} and \eqref{eq:spectral} we see that a KMS state is passive in the sense
that, for any Hermitian operator $\mathcal{O}$, its spectral density for any frequency obeys
\begin{equation}
\omega\,\rho_\mathcal{O}(\omega) \geq 0. 
\label{eq:passivity}
\end{equation}
Hence, for a positive-frequency perturbation, $\rho_{\mathcal O}(\omega)\geq0$, and the dissipative part of the retarded response has the absorptive sign which is also evident  from \eqref{eq:spectral_Im}.
 
The positive semidefiniteness of $\Gamma $ \eqref{eq:Gamma_positive} implies a non-positive imaginary part of the retarded self-energy \eqref{eq:SelfEnergy_decomp}
\begin{equation}
-\mathrm{Im}\,\Sigma^R(\omega) \geq 0, 
\label{eq:absorptive}
\end{equation} 
 which has the absorptive sign. 
KMS passivity therefore provides the finite-temperature QFT interpretation of the decay width, where the open worldsheet channel absorbs energy rather than acting as a gain medium. Consequently, in a stable passive state, a leaking photon ring resonance appears in the retarded response as a pole in the lower half of the complex frequency plane. A pole with the opposite sign would correspond to negative absorption and therefore to amplification by a non-passive state, rather than to QNM leakage.
 
%=====================================================================
\section{Half-density realization of the projected escape channel}
\label{section:halfd}
%=====================================================================
 \subsection{Thermal derivation of complex frequencies}
 
 Motivated by the standard thermal two-point function of a one-dimensional conformal kernel of weight $h$,  we introduce a useful effective description of the near-ring resonance channel in terms of a one-dimensional thermal Rindler operator $\mathcal O_h $ of weight $h$, with Euclidean two-point function 
\begin{equation}
G_{\rm E}(\tau_{\rm E}) = C_h\!\left(\frac{\pi T_{\rm ind}}{\sin(\pi T_{\rm ind}\tau_{\rm E})}\right)^{\!2h}.
\label{eq:GE}
\end{equation}
We use this conformal form as an effective kernel characterized by its dilation weight and thermal covariance.  The full gauge-fixed worldsheet theory is not assumed as a one-dimensional conformal field theory. After analytic continuation $\tau_{\rm E} = i\tau+\epsilon$,  and absorbing the overall branch-dependent phase into the normalization $C_h$, we select the boundary value corresponding to the $G^>$ ordering and obtain
\begin{equation}
G^>(\tau) = C_h\!\left(\frac{\pi T_{\rm ind}}{\sinh[\pi T_{\rm ind}(\tau-i\epsilon)]}\right)^{\!2h}.
\label{eq:Wightman_cont}
\end{equation}
To obtain the pole locations let us make the positive time expansion for $\tau>0$,
\begin{align}
\left(\frac{\pi T_{\rm ind}}{\sinh(\pi T_{\rm ind}\tau)}\right)^{\!2h}
&= (2\pi T_{\rm ind})^{2h}\,e^{-2\pi T_{\rm ind}h\tau}\!\left(1-e^{-2\pi T_{\rm ind}\tau}\right)^{-2h}
  \label{eq:expansion1}\\
&= (2\pi T_{\rm ind})^{2h}\sum_{n=0}^\infty \frac{(2h)_n}{n!}\, e^{-2\pi T_{\rm ind}(n+h)\tau},
\label{eq:expansion2}
\end{align}
where $(a)_n$ is the Pochhammer symbol.

The Wightman kernel is not itself the retarded correlator. For the pole analysis, we assume that the nonlocal positive-time part of the commutator has the same decay exponents as the thermal Wightman kernel, although its coefficients may differ. Thus, after analytic continuation, the positive time Lorentzian kernel \eqref{eq:Wightman_cont}, equivalently its expansion \eqref{eq:expansion2}, is Fourier transformed to determine the pole structure as
\begin{align}
{\tilde G}^R_{(h)}(\omega_b)
 &\sim -i\int_0^\infty d\tau\,e^{i\omega_b\tau}\left(\frac{\pi T_{\rm ind}}{\sinh(\pi T_{\rm ind}\tau)}\right)^{2h} + \text{contact terms} \label{eq:GR_h_int}\\
&\sim -i(2\pi T_{\rm ind})^{2h}\sum_{n=0}^\infty\frac{(2h)_n}{n!}\, \frac{1}{2\pi T_{\rm ind}(n+h)-i\omega_b} + \text{contact terms},
\label{eq:GR_h_poles}
\end{align}
and therefore the poles are at
\begin{equation}
\omega_{b,n} = -i\,2\pi T_{\rm ind}(n+h) = -i\kappa_{\rm ind}(n+h),\qquad n=0,1,2,\ldots
 \label{eq:poles_b}
\end{equation}
Here $\omega_b$ is the frequency measured with respect to the co-rotating photon ring time \eqref{eq:corot-frequency},
\begin{equation}
\omega_b =\omega^{\rm cr}= \omega - m\Omega_{\rm orb},
\label{eq:omegab}
\end{equation}
so that
\begin{equation}
\omega_{mn} = m\Omega_{\rm orb} - i\lambda_L(n+h).
 \label{eq:QNM_h}
\end{equation}
The azimuthal quantum number $m$ is not a dynamical quantum number of the worldsheet theory.
The probe string describes the local near-ring dynamics in the frame co-rotating with the
reference photon orbit.  The thermal response of the worldsheet therefore determines the shifted
frequency $\omega_b$ in Eq.~\eqref{eq:QNM_h}, and in particular the imaginary part and the
overtone spacing controlled by $\lambda_L$.  The integer $m$ enters only when translating back
from the co-rotating worldsheet frequency $\omega_b=\omega^{\rm cr}$ to the asymptotic observer frequency
$\omega$ (measured with respect to the asymptotic time~$t$), producing the real part
$m\Omega_{\rm orb}$ in the eikonal spectrum.

 \subsection{Thermal monodromy}
 \noindent
An inspection of Eq.~\eqref{eq:QNM_h} shows that the eikonal QNM spectrum
in Eq. ~\eqref{eq:qnm} is reproduced by
the value  $ h = \frac{1}{2}$, which, however, cannot be determined by the KMS relation.  In other words, the KMS relation
alone fixes the thermal scale $2\pi T_{\rm ind}=\kappa_{\rm ind}$ and the analytic strip.
 The offset is the operator data.  In the next subsection, we will determine the value $h=1/2$ by looking at the specific resonance channel that represents radial escape from the photon ring.  

For this value of $h$, a useful subtlety arises in the thermal monodromy of the kernel. Under a shift by one Euclidean Rindler period, $\tau\to\tau+i\beta$, with $\beta_{\rm ind}=1/T_{\rm ind}$, one has
\begin{equation}
\sinh[\pi T_{\rm ind}(\tau+i\beta_{\rm ind}-i\epsilon)] = -\sinh[\pi T_{\rm ind}(\tau-i\epsilon)].
\label{eq:monodromy}
\end{equation}
Therefore, a kernel of weight $h$ acquires the phase
\begin{equation}
 G^>_h(\tau+i\beta_{\rm ind}) = e^{-2\pi i h}\,G^>_h(\tau),
\label{eq:monodromy2}
\end{equation}
up to the branch prescription.  In particular, for $h=1/2$ the projected escape kernel acquires a minus sign under thermal monodromy around the Euclidean Rindler circle,
\begin{equation}
G^>_{1/2}(\tau+i\beta_{\rm ind}) = -G^>_{1/2}(\tau).
\label{eq:antiperiodic}
\end{equation}
This is a fermion-like thermal monodromy, but it does not mean that the escape operator is a
fundamental fermionic worldsheet field.  As shown below, the sign originates from the half-density monodromy, equivalently from the square-root Jacobian required by the unitary boost representation.  This is similar in spirit to the statistics-inversion phenomena familiar from
Unruh-type responses and to the appearance of fermion-like thermal structures in Rindler and
inverted-oscillator descriptions of QNM
systems~\cite{Takagi:1986kn,Arrechea:2021szl,Hegde:2018xub}.  For the pole locations
derived below, however, only the positive-time decay factors are needed.

\subsection{The value of $h$: the eikonal QNM spectrum}

 Below we give a projected worldsheet
derivation in which the single unstable Penrose-limit direction gives a half-density
representation of Rindler dilations, and therefore an effective radial weight $h=1/2$.
 
The relevant object in the following discussion is not assumed to be a fundamental local Nambu–Goto field. Rather, it is a projected operator associated with the effective near-ring escape channel. In the static gauge introduced in \eqref{eq:staticgauge}, the string is extended along the stable transverse Penrose direction, and this embedding induces the Rindler worldsheet metric \eqref{eq:Rindler_ws}. 

The unstable Penrose direction $x^1$ remains an ordinary local worldsheet fluctuation whose zero mode gives the inverted-oscillator instability described in Section~\ref{sec:tidal-fluc}. The projected escape operator introduced below should be understood as an effective response-theory representation of this same near-ring leakage channel, expressed in terms of the dilation flow of the induced Rindler geometry. The null geodesics of the induced Rindler geometry \eqref{eq:Rindler_ws} exhibit the universal dilation structure generated by worldsheet-time translations:
\begin{equation}
\frac{d\sigma}{d\tau} = \pm\kappa_{\rm ind}\sigma,\qquad \sigma(\tau) = \sigma_0\,e^{\pm\kappa_{\rm ind}\tau}.
  \label{eq:null_geod}
\end{equation}
In Rindler null coordinates $X^\pm$ introduced in \eqref{eq:lightcone}, a translation $\tau\to\tau+a$ acts as the boost \eqref{eq:boost}. To connect the boost description with the microscopic unstable mode, we project $x^1$ onto its zero worldsheet-momentum component $\eta_0$, whose Hamiltonian is the inverted oscillator \eqref{eq:io-H}. It is useful to introduce the phase-space combinations
\begin{equation}
\chi_+=\frac{p_\eta+\lambda_L\eta_0}{\sqrt{2\lambda_L}},
\qquad
\chi_-=\frac{p_\eta-\lambda_L\eta_0}{\sqrt{2\lambda_L}},
\end{equation}
where we know that $\lambda_L=\kappa_{\rm ind}$ from \eqref{eq:Rindler_ws}. 
Their evolution under the Hamiltonian \eqref{eq:io-H} is
\begin{equation}
\chi_+(\tau)=e^{\lambda_L\tau}\chi_+(0),
\qquad
\chi_-(\tau)=e^{-\lambda_L\tau}\chi_-(0).
\end{equation}
Upon quantization these variables satisfy
\begin{equation}
[\chi_+,\chi_-]=i,
\end{equation}
and the symmetrically ordered inverted-oscillator Hamiltonian becomes
\begin{equation}
H_{\rm ws}^{(0)}=\frac{\kappa_{\rm ind}}{2}\left(\chi_+\chi_-+\chi_-\chi_+\right).
\label{eq:IHO_dilation}
\end{equation}
We identify $\chi\equiv\chi_+$ with the expanding phase-space variable of the unstable zero mode and use it as the reduced coordinate for one outgoing escape branch. We restrict to one expanding branch, $\chi>0$, as an effective description of one outgoing escape channel. The full two-arm Gamow problem requires combining the corresponding branches through the usual analytic continuation, and that global construction is not needed for the local pole-spacing argument developed here. The projected operator $\mathcal O_{\rm esc}$ can then be understood as an effective response operator associated with this outgoing unstable channel after the remaining worldsheet and exterior degrees of freedom have been integrated out. The Rindler geometry supplies the universal boost structure and temperature, while the unstable zero mode supplies the microscopic escape degree of freedom.

Classically, the Rindler time evolution of this coordinate is hyperbolic,
\begin{equation}
\chi(\tau) = e^{\kappa_{\rm ind}\tau}\chi(0),
\label{eq:chi_evol}
\end{equation}
and the corresponding radial Hilbert space may be taken to be
$\mathcal{H}_{\rm rad}=L^2(\mathbb{R}_+,d\chi)$. Here we are using the known relation that the Lyapunov exponent is equal to the induced surface gravity \eqref{eq:Rindler_ws}.  

The essential observation now is that the
action of the Rindler boost on this Hilbert space is fixed by unitarity. Let us study how the wavefunctions transform under this dilation so that the quantum evolution is unitary.  Under a shift in Rindler time, the classical map
\begin{equation}
\chi\longrightarrow\chi'=e^{\kappa_{\rm ind} a}\chi
\label{eq:boost_classical}
\end{equation}
is represented on wavefunctions by
\begin{equation}
\bigl(U(a)\psi\bigr)(\chi) = e^{-\kappa_{\rm ind} a/2}\,\psi(e^{-\kappa_{\rm ind} a}\chi),
\label{eq:boost_unitary}
\end{equation}
where  $U(a)$ is the dilation operator and the factor $e^{-\kappa_{\rm ind} a/2}$  is fixed uniquely by unitarity. It comes from the Jacobian and the conservation of the radial norm
\begin{equation}
\int d\chi\,|(U(a)\psi)(\chi)|^2 = \int d\chi\,e^{-\kappa_{\rm ind}a}\,|\psi(e^{-\kappa_{\rm ind}a}\chi)|^2 = \int d\chi'\,|\psi(\chi')|^2.
\label{eq:norm_conserved}
\end{equation}
Thus the radial wavefunction transforms as a half-density under the Rindler boost.
The generator of this boost action is
\begin{equation}
K_\chi = -i\!\left(\chi\partial_\chi + \frac{1}{2}\right), \label{eq:Kchi}
\end{equation}
up to the sign convention for the direction of the boost.  
On the expanding branch, $\chi=\chi_+>0$, we may represent
$\chi_-=-i\partial_\chi$, and   Eq.~\eqref{eq:IHO_dilation} then gives
\begin{equation}
H_{\rm ws}^{(0)}=-i\kappa_{\rm ind}
\left(\chi\partial_\chi+\frac12\right)=\kappa_{\rm ind}K_\chi.
\label{eq:IHO_Kchi}
\end{equation}
Thus the inverted oscillator realizes exactly the unitary
half-line dilation representation.

The same unitary dilation representation occurs for geometric boosts
on a positive Rindler null half-line. On $L^2(\mathbb R_+,dX^+)$ the corresponding
one-particle generator is
\begin{equation}
K_+^{(1)}=-i\left(X^+\partial_{X^+}+\frac12\right),
\end{equation}
and under the identification $X^+=\ell\chi$, with arbitrary $\ell>0$, the
unitary map
\begin{equation}
(W\psi)(X^+)=\ell^{-1/2}\psi(X^+/\ell)
\end{equation}
satisfies
\begin{equation}
WK_\chi W^{-1}=K_+^{(1)}.
\end{equation}
This proves the unitary equivalence between the reduced
inverted-oscillator representation and the one-particle half-line
representation of the geometric boost.  This is a statement at the
level of the reduced boost representation, and it does not identify
$K_\chi$ with the complete modular generator $K_{\rm BW}$.  Rather,
$K_\chi$ should be interpreted as the dimensionless boost generator
acting on the projected escape channel.

The corresponding dimensionful generator of $\tau$ evolution in this reduced representation is given by \eqref{eq:IHO_Kchi}, 
$H_{\rm ws}^{(0)}=\kappa_{\rm ind}K_\chi.$
The constant $1/2$ in $K_\chi$ is therefore a kinematic consequence of
the unitary dilation representation on
$L^2(\mathbb R_+,d\chi)$.  It is the half-density correction to the
classical dilation generator $\chi\partial_\chi$.

Unitarity thus fixes the half-density transformation law of the
reduced escape wavefunction.  To construct an effective response
operator, we represent $\mathcal O_{\rm esc}$ with the same
half-density covariance.  Within this effective realization, the
projected escape channel carries the weight
\begin{equation}
h=\frac12.
\label{eq:h=1/2}
\end{equation}
This assignment concerns the projected escape amplitude and should not
be interpreted as the conformal weight of the local Nambu--Goto
fluctuation.  Within this effective half-density realization,
substituting \eqref{eq:h=1/2} into \eqref{eq:QNM_h} gives
\begin{equation}
\omega_{mn}=m\Omega_{\rm orb}-i\lambda_L\left(n+\frac12\right),
\label{eq:QNM_h3}
\end{equation}
which reproduces the eikonal QNM spectrum already obtained from the
outgoing inverted-oscillator resonances in
\eqref{eq:qnm}.

It is also possible to represent the half-density covariance through
an effective two-point kernel on the reduced escape coordinate.  When
pulled back to the boost orbit, this kernel takes the characteristic
form proportional to $1/\sinh(\pi T_{\rm ind}\tau)$.  This construction exhibits the same
Rindler scale and half-density covariance as the microscopic
inverted-oscillator problem, but it is not used as an independent
derivation of the retarded QNM poles.  The construction is presented
in Appendix~\ref{twopoint}.
 
%=====================================================================
\section{Conclusions}
\label{section:conclusions}
%=====================================================================
 
We have shown that the eikonal QNM spectrum admits a natural interpretation as the causal thermal response of a probe string worldsheet near the photon ring. The Penrose-limit geometry induces a Rindler worldsheet whose surface gravity equals the photon ring Lyapunov exponent. The regular worldsheet state is therefore a KMS state with temperature $T_{\rm ind}$, providing a thermal description of the near-ring dynamics. In this sense, the photon ring
Lyapunov exponent is reinterpreted as a temperature scale in a finite-temperature worldsheet
quantum field theory. 

The induced Rindler worldsheet provides only the thermal background. The QNM itself originates from the unstable transverse worldsheet fluctuation. Its zero worldsheet-momentum component is an inverted harmonic oscillator governed by the photon ring Lyapunov exponent. In the co-rotating frame its outgoing Gamow resonances reproduce the eikonal QNM spectrum, thereby identifying the microscopic worldsheet excitation that underlies the effective thermal response.

The damping of QNMs follows from treating the near-ring region as an open subsystem. After projecting onto the escape channel, the retarded response acquires a self-energy whose imaginary part describes leakage into the asymptotic exterior and the black hole horizon. Retarded causality fixes the analyticity of the response, spectral positivity ensures the absorptive sign of the decay width, while KMS passivity gives this sign its thermal interpretation. Together, these ingredients explain why physical QNMs appear as poles in the lower half of the complex-frequency plane.

 The half-integer offset has a different origin. KMS determines the thermal scale, but does not fix the value of the offset. We have shown that the projected escape coordinate carries the unitary half-density representation of Rindler dilations. The resulting boost weight $h=\tfrac12$ reproduces the characteristic half-integer structure of the eikonal QNM spectrum without assigning conformal weight 1/2 to the underlying local Nambu–Goto field.

Thus the final picture is the following. The Penrose-limit geometry induces a Rindler worldsheet on the probe string whose surface gravity equals the photon ring Lyapunov exponent. The unstable transverse worldsheet fluctuation provides the microscopic degree of freedom whose zero mode reproduces the eikonal QNM spectrum. The regular worldsheet state supplies KMS thermality, while projection onto the escape channel turns the near-ring sector into an open subsystem whose retarded response acquires an absorptive self-energy. Finally, the unitary boost representation of the projected escape coordinate fixes the half-density weight responsible for the half-integer overtone shift. Together these ingredients provide a thermal worldsheet interpretation of the eikonal QNM  spectrum.  The worldsheet fluctuation analysis and the thermal response analysis therefore provide complementary descriptions of the same eikonal QNM spectrum. The former identifies the microscopic unstable degree of freedom, while the latter explains the thermal and causal structure of its resonance poles. 
 
It would be interesting to extend the present construction beyond the strict eikonal
limit.  In the leading Penrose-limit description, the near-ring geometry is captured by the
quadratic Brinkmann profile, and the worldsheet dynamics reduces to the induced Rindler problem
studied above.  Subleading corrections in the eikonal expansion should arise from higher orders
in the near-geodesic expansion of the ambient metric, as well as from the nonlinear terms in
the Nambu--Goto action.  In the static gauge $x^2=\sigma$, such corrections are expected to generate interactions between the local Nambu–Goto fluctuations and the projected escape channel associated with the unstable Penrose zero mode. Such corrections may therefore modify not only the  effective retarded self-energy of the projected escape channel but also the coupling between the microscopic unstable worldsheet fluctuation and the reduced thermal response.
 
A useful question is whether these corrections merely shift the real and imaginary parts of the
QNM poles through corrections to $\Sigma^R(\omega)$, or whether they also modify the effective
half-density weight of the projected escape channel.  In the strict eikonal limit, this weight
is fixed kinematically by the unitary boost action on $L^2(\mathbb{R}_+,d\chi)$ and gives
$h=1/2$.  Understanding whether the half-density weight is protected or receives controlled finite-eikonal corrections would sharpen the worldsheet description of subleading QNM effects. It would also clarify how the microscopic unstable worldsheet mode is encoded in the projected thermal response beyond the strict eikonal limit.

Finally, an interesting question is whether our results are connected to the recently remarked link between  the thermal operator product expansion and QNMs in large-$N$ thermal conformal field theories \cite{Arnaudo:2026axe,Barrat:2026jfg}.

\section*{Acknowledgements}
D.G.  acknowledges support from  the National Science and Technology Council (NSTC) of Taiwan with the Young Scholar Columbus Fellowship grant 114-2636-M-110-004 and 115-2112-M-110-010.  F.Q.'s research is funded by Tamkeen under the research grant to NYUAD ADHPG-AD457.  A.R.  acknowledges support from the Swiss National Science Foundation (project number CRSII5\_213497).

%=====================================================================
\appendix
\section{The two-point kernel}
\label{twopoint}

We now introduce a convenient local two-point kernel that realizes the half-density
transformation law derived above. This kernel is an effective representation of the
projected escape amplitude, not a uniquely determined correlator of the microscopic
worldsheet theory. Under a rescaling
\begin{equation}
\chi \longrightarrow \chi'=\Lambda\chi,
\end{equation}
a half-density transforms with the square root of the inverse Jacobian,
\begin{equation}
\mathcal O'_\chi(\chi')=\left(\frac{d\chi}{d\chi'}\right)^{1/2}
\mathcal O_\chi(\chi)=\Lambda^{-1/2}\mathcal O_\chi(\chi).
\end{equation}

Locally away from the boundary at $\chi=0$, a convenient translation-invariant
short-distance kernel with this covariance is
\begin{equation}
\widetilde{K}_\chi(\chi_1,\chi_2)
\equiv\left\langle\mathcal O_\chi(\chi_1)\mathcal O_\chi(\chi_2)\right\rangle
=\frac{\mathcal N}{\chi_1-\chi_2-i0^+},
\label{eq:kernel}
\end{equation}
up to contact terms and an overall normalisation. Under
$\chi_i\to\Lambda\chi_i$, it scales as
\begin{equation}
\widetilde{K}_\chi(\Lambda\chi_1,\Lambda\chi_2)=\Lambda^{-1}\widetilde{K}_\chi(\chi_1,\chi_2),
\label{eq:kernel_covariance}
\end{equation}
as expected for a pair of half-density insertions. The covariance is understood
in the $i0^+$ boundary-value sense, with the regulator scaled consistently before
taking the limit. Equation~\eqref{eq:kernel} should therefore not be interpreted
as the correlator of a fundamental local Nambu--Goto scalar. It should instead be
regarded as a convenient effective short-distance kernel for the reduced
one-dimensional escape problem.

The corresponding Rindler-time escape operator is obtained by pulling this
half-density back to the boost orbit $\chi=\chi(\tau)$:
\begin{equation}
\mathcal O_{\rm esc}(\tau)=\left(\frac{d\chi}{d\tau}\right)^{1/2}\mathcal O_\chi(\chi(\tau)).
\label{eq:Oesc}
\end{equation}
The Jacobian factor is the standard pullback of a half-density from the coordinate
$\chi$ to the Rindler-time orbit. Therefore,
\begin{equation}
G^>_{\rm esc}(\tau)
\equiv\left\langle\mathcal O_{\rm esc}(\tau)\mathcal O_{\rm esc}(0)
\right\rangle
=\mathcal N\frac{\bigl[\dot\chi(\tau)\dot\chi(0)\bigr]^{1/2}}{\chi(\tau)-\chi(0)-i0^+}.
\label{eq:Gesc}
\end{equation}
Using
$\chi(\tau)=\chi_0 e^{\kappa_{\rm ind}\tau}$,
we first obtain
\begin{equation}
G^>_{\rm esc}(\tau)=\mathcal N
\frac{\kappa_{\rm ind}\chi_0 e^{\kappa_{\rm ind}\tau/2}}{\chi_0\bigl(e^{\kappa_{\rm ind}\tau}-1\bigr)-i0^+}.
\label{eq:Gesc_intermediate}
\end{equation}
Mapping the coordinate-space boundary prescription to the corresponding
time-domain $i0^+$ boundary value, and absorbing constant factors into
$\mathcal N$, this becomes
\begin{equation}
G^>_{\rm esc}(\tau)=\mathcal N\frac{\kappa_{\rm ind}/2}{\sinh\!\left[\frac{\kappa_{\rm ind}}{2}(\tau-i0^+)\right]},
\label{eq:Gesc2}
\end{equation}
or, equivalently, using \eqref{eq:Tind},
\begin{equation}
G^>_{\rm esc}(\tau)=\mathcal N\frac{\pi T_{\rm ind}}{\sinh\!\left[\pi T_{\rm ind}(\tau-i0^+)\right]}.
\label{eq:Gesc3}
\end{equation}
With the boundary prescription chosen above, this expression defines the
$G^>$ boundary value of the projected escape kernel. The power one in the
denominator follows from the half-density boost representation, rather than
from an independently assigned effective conformal dimension for a local
worldsheet operator.

The QNM frequencies are poles of the causal retarded response, 
\begin{equation}
G^R_{\rm esc}(\tau)=-i\theta(\tau)\left\langle\left[\mathcal O_{\rm esc}(\tau),\mathcal O_{\rm esc}(0)\right]\right\rangle,
\label{eq:GResc2}
\end{equation}
and not of the
unordered Wightman correlator itself. To show how this effective kernel can reproduce the microscopic pole
tower, we now make a sensible  additional response-theory assumption. We assume
that the nonlocal positive-time part of the retarded commutator has the
same decay exponents as \eqref{eq:expansion_pos}, although its
coefficients may differ. This assumption is not implied by the KMS
condition and is not derived here from the microscopic Nambu--Goto
theory. Under this assumption, the nonlocal part of the retarded kernel
may be written as
\begin{equation}
G^R_{\rm esc}(\tau)\simeq-i\theta(\tau)\sum_{n=0}^{\infty}a_n e^{-\kappa_{\rm ind}(n+1/2)\tau}+\text{local terms},
\label{eq:GResc_pos}
\end{equation}
where the coefficients $a_n$ absorb the overall normalisation and the relative
commutator weights. Their precise values do not affect the pole positions. In deriving Eq. \eqref{eq:GResc_pos}, the relation
\begin{equation}
\frac{1}{\sinh(\kappa_{\rm ind}\tau/2)}=2\sum_{n=0}^{\infty}e^{-\kappa_{\rm ind}(n+1/2)\tau},
\label{eq:expansion_pos}
\end{equation}
for $\tau>0$ has been used.
Fourier transforming \eqref{eq:GResc_pos} with respect to the co-rotating frequency $\omega_b$, we
obtain
\begin{align}
{\tilde G}^R_{\rm esc}(\omega_b)
&\simeq-i\sum_{n=0}^{\infty}a_n\int_0^\infty d\tau\,e^{i\omega_b\tau}e^{-\kappa_{\rm ind}(n+1/2)\tau}
\nonumber\\
&=-i\sum_{n=0}^{\infty}a_n \frac{1}{\kappa_{\rm ind}\left(n+\frac12\right)-i\omega_b}.
\label{eq:GResc_poles}
\end{align}
Equivalently, defining $c_n=-ia_n$, we can express $G^R_{\rm esc}$ as
\begin{equation}
{\tilde G}^R_{\rm esc}(\omega_b)\simeq \sum_{n=0}^{\infty}\frac{c_n}{\kappa_{\rm ind}\left(n+\frac12\right)-i\omega_b}+\text{contact terms},
\label{eq:GResc_sum}
\end{equation}
where the contact terms arise from local in time contributions to the retarded
function, and they produce polynomial terms in frequency space. Therefore they do
not affect the pole positions which  satisfy
\begin{equation}
\kappa_{\rm ind}\left(n+\frac12\right)-i\omega_b=0,
\label{eq:pole_eq}
\end{equation}
Hence, the frequencies are given by
\begin{equation}
\omega_{b,n}=-i\kappa_{\rm ind}\left(n+\frac12\right), \qquad n=0,1,2,\ldots, 
\label{eq:poles_final}
\end{equation}
which, after restoring the orbital contribution, turn out to be
\begin{equation}
\omega_{mn}=m\Omega_{\rm orb}-i\lambda_L\left(n+\frac12\right),\qquad n=0,1,2,\ldots,
\label{eq:QNM_final}
\end{equation}
using \eqref{eq:Rindler_ws} and \eqref{eq:omegab}.
Thus, under the response-theory assumption stated above, the effective
half-density kernel reproduces the QNM pole tower already obtained from
the outgoing inverted-oscillator resonances. This is a consistency
check of the effective description, rather than an independent
derivation of the pole spectrum.

The KMS property of the Rindler worldsheet fixes the temperature and
the thermal analytic strip,
$\kappa_{\rm ind}=2\pi T_{\rm ind}$, but it does not by itself imply an
evenly spaced retarded pole tower. In the present effective
construction, the spacing follows from the $1/\sinh$ kernel together
with the response-theory assumption stated above. The half-integer
offset is encoded by the half-density transformation law of the
projected outgoing phase-space coordinate associated with the unstable
Penrose zero mode.

The half-density statement concerns a projected effective channel rather than
the full local Nambu--Goto field. The unstable fluctuation
$x^1(\tau,\sigma)$ has its own quadratic worldsheet dynamics, and its zero
worldsheet-momentum component $\eta_0(\tau)$ provides the microscopic
inverted-oscillator degree of freedom associated with near-ring escape. The
reduced coordinate $\chi$ is constructed from the outgoing component of this
zero mode, and $\mathcal O_{\rm esc}$ is the corresponding effective response
operator after projection onto the escape channel.

In the static gauge $x^2=\sigma$, the stable Penrose direction supplies the
radial coordinate of the induced Rindler worldsheet, while the orthogonal
unstable fluctuation supplies the microscopic escape degree of freedom. Thus,
$h=1/2$ is the effective boost weight of the projected outgoing zero-mode
amplitude. It follows from the unitary dilation representation on the reduced
$\chi$-space, rather than from assigning conformal weight $1/2$ to the local
field $x^1(\tau,\sigma)$.
Combining the above, the effective half-density kernel provides
a response-theory realization of the same pole tower obtained from the
microscopic inverted oscillator. The pole spectrum itself remains fixed
by the outgoing Gamow condition, while the present construction shows
how the same thermal scale and half-density structure can be encoded in
an effective projected kernel.

\bibliographystyle{JHEP}
\bibliography{darftjhep_latest_version}

\end{document}